\documentstyle[a4,12pt]{article}
\renewcommand{\baselinestretch}{1.5}

\begin{document}
\parskip=5pt plus 1pt minus 1pt

\setcounter{page}{0}

\begin{flushright}
{LMU-2/94}\\
{March 1994}
\end{flushright}

\vspace{0.5cm}

\begin{center}
{\Large\bf On Test of $CPT$ Symmetry in $CP$-violating $B$ Decays}
\end{center}

\vspace{0.5cm}

\begin{center}
{\bf Zhi-zhong Xing}\footnote{{\it Alexander von Humboldt} Research Fellow; $~$
E-mail: Xing$@$hep.physik.uni-muenchen.de}
\end{center}

\begin{center}
{\sl Sektion Physik, Ludwig-Maximilians-Universit${\sl\ddot a}$t M${\sl\ddot
u}$nchen}\\
{\sl Theresienstra$\beta$e 37, D-80333 Munich, Germany}
\end{center}

\vspace{1cm}

\begin{center}
{\bf Abstract}
\end{center}

\small
	Considering that the existing experimental limit for $CPT$ violation is
still poor, we explore various possible ways to test $CPT$ symmetry in
$CP$-violating
$B$ decays at $e^{+}e^{-}$ $B$ factories and hadron machines.
We find that it is difficult to distinguish between the effect of direct
$CP$ violation and that of $CPT$ violation in the time-integrated
measurements of neutral $B$ decays to $CP$ eigenstates such as
$\psi K_{S}$ and $\pi^{+}\pi^{-}$. Instead, a cleaner signal of small
$CPT$ violation may appear in the
time-integrated $CP$ asymmetries for a few non-$CP$-eigenstate channels,
e.g., $B^{0}_{d}/\bar{B}^{0}_{d}\rightarrow D^{\pm}\pi^{\mp}$
and $\stackrel{(-)}{D}$$^{(*)0}K_{S}$. The time-dependent measurements of
$CP$ asymmetries are available, in principle, to limit the size of
$CPT$-violating effects
in neutral $B$-meson decays.

\normalsize

\vspace{2cm}


\newpage

\begin{flushleft}
{\Large\bf I. $~$ Introduction}
\end{flushleft}

	Tests of the fundamental symmetries and conservation laws
have been an important topic in particle physics.
Recently the necessity of testing $CPT$ symmetry has been emphasized
by several authors [1-3]. On the experimental side, the existing evidence
for $CPT$ invariance is rather poor. The limit for the strength of
$CPT$-violating interaction in the $K^{0}-\bar{K}^{0}$ system is
about $10\%$ of that of $CP$-violating interaction [2-4].
This means that $CPT$ symmetry is tested only at the $10\%$ level.
On the theoretical
side, the universality of $CPT$ theorem is questionable since it is
proved in local renormalizable field theory with a heavy use of the
properties of asymptotic states [5]. This proof may not be applicable
for QCD, where quarks and gluons are confined other than in a set of
asymptotic states [3].

	The simplest test of $CPT$ invariance is to examine the
equality of the masses and lifetimes of a particle and its antiparticle.
Beyond the $K$-meson system, Kobayashi and Sanda
are the first to suggest some ways to check $CPT$ symmetry in $B$-meson decays
at the future $B$ factories [3]. With the assumption that the
amplitudes for semileptonic decays satisfy the $\Delta Q=\Delta B$ rule
and $CPT$ invariance, they relaxed $CPT$ symmetry for the mass matrix
of the $B_{d}$ mesons. They found that both
$B^{0}_{d}-\bar{B}^{0}_{d}$ mixing and $CP$ asymmetries in neutral $B$
decays will get modified if $CPT$ violation is present. A key point of their
work is that $CP$ asymmetries in nonleptonic $B$ decays such as
$B^{0}_{d}/\bar{B}^{0}_{d}\rightarrow \psi K_{S}$ may be more
sensitive to the presence of small $CPT$-violating effects than the
dileptonic decay rates of $B^{0}_{d}\bar{B}^{0}_{d}$ pairs.

	In this work, we shall make an
instructive analysis of the effects of $CPT$ violation
on $CP$-violating asymmetries in neutral $B$-meson decays.
Both time-dependent and time-integrated
$CP$ asymmetries are calculated to
meet various possible measurements at $e^{+}e^{-}$
$B$ factories and hadron machines. We suggest several ways for distinguishing
$CPT$ violation from direct $CP$ violation in $B$ decay amplitudes
and indirect $CP$ violation via interference between decay and
mixing.
We show that it is difficult to extract the $CPT$-violating information
from the time-integrated measurements of neutral $B$ decays to $CP$ eigenstates
such as
$B^{0}_{d}/\bar{B}^{0}_{d}\rightarrow \psi K_{S}$ and $\pi^{+}\pi^{-}$.
Instead, cleaner signals of small $CPT$
violation may appear in the time-integrated $CP$ asymmetries of some
non-$CP$-eigenstate channels, e.g., $B^{0}_{d}/\bar{B}^{0}_{d}\rightarrow
D^{\pm}\pi^{\mp}$ and $\stackrel{(-)}{D}$$^{(*)0}K_{S}$.
Measurements of the time development of $B^{0}_{d}$ versus $\bar{B}^{0}_{d}$
decays are available, in principle, to limit
the size of $CPT$-violating effects on $CP$ asymmetries. \\

\begin{flushleft}
{\Large\bf II. $~$ Decay probabilities}
\end{flushleft}

	We begin with the mass eigenstates of the two $B_{d}$ mesons [6]:
\begin{eqnarray}
|B_{1}> & = & \frac{1}{\sqrt{|p_{1}|^{2}+|q_{1}|^{2}}}\left (
p_{1}|B^{0}_{d}> + q_{1}|\bar{B}^{0}_{d}> \right ) \; , \nonumber \\
|B_{2}> & = & \frac{1}{\sqrt{|p_{2}|^{2}+|q_{2}|^{2}}}\left (
p_{2}|B^{0}_{d}> - q_{2}|\bar{B}^{0}_{d}> \right ) \; ,
\end{eqnarray}
where $p_{1,2}$ and $q_{1,2}$ are parameters of the $B^{0}_{d}
\bar{B}^{0}_{d}$ mass matrix elements. For convenience, the ratios of
$q_{1,2}$ to $p_{1,2}$ can be further written as
\begin{equation}
\frac{q_{1}}{p_{1}}\; =\; e^{i\phi}\tan\frac{\theta}{2} \; ,
\;\;\;\;\; \frac{q_{2}}{p_{2}}\; =\; e^{i\phi}\cot\frac{\theta}{2} \; ,
\end{equation}
where $\theta$ and $\phi$ are generally complex [6]. $CPT$ symmetry
requires $q_{1}/p_{1}=q_{2}/p_{2}=e^{i\phi}$ or ${\cal S}\equiv
\cot\theta =0$, and $CP$ invariance requires that both ${\cal S}=0$
and $\phi =0$ hold. Furthermore, the proper-time evolution of an
initially ($t=0$) pure $B^{0}_{d}$ or $\bar{B}^{0}_{d}$ is given by
\begin{eqnarray}
|{{B}^{0}_{d}}(t)> & = & e^{-\left (im+\frac{\Gamma}{2}\right )
t}\left [g_{+}(t)|B^{0}_{d}> + \tilde{g}_{+}(t)|\bar{B}^{0}_{d}>
\right ] \; , \nonumber \\
|\bar{B}^{0}_{d}(t)> & = & e^{-\left (im+\frac{\Gamma}{2}\right )
t}\left [\tilde{g}_{-}(t)|B^{0}_{d}> + g_{-}(t)|\bar{B}^{0}_{d}>
\right ] \; ,
\end{eqnarray}
where
\begin{eqnarray}
g_{\pm}(t) & = & \cos^{2}\frac{\theta}{2}e^{\pm\left (i\Delta m-\frac{\Delta
\Gamma}{2}\right )\frac{t}{2}}
+ \sin^{2}\frac{\theta}{2}e^{\mp\left (
i\Delta m-\frac{\Delta \Gamma}{2}\right )\frac{t}{2}}
\; , \nonumber\\
\tilde{g}_{\pm}(t) & = & \sin\frac{\theta}{2}\cos\frac{\theta}{2}
\left [e^{\left (i\Delta m-\frac{\Delta \Gamma}{2}\right )\frac{t}{2}}
-e^{-\left (i\Delta m-\frac{\Delta \Gamma}{2}\right )\frac{t}{2}}\right ]
e^{\pm i\phi} \; .
\end{eqnarray}
In Eqs.(3) and (4) we have defined $m\equiv (m_{1}+m_{2})/2$,
$\Gamma\equiv (\Gamma_{1}+\Gamma_{2})/2$, $\Delta m \equiv m_{2}-m_{1}$,
and $\Delta \Gamma \equiv \Gamma_{1}-\Gamma_{2}$, where $m_{1,2}$ and
$\Gamma_{1,2}$ are the mass and width of $B_{1,2}$ .

	Concentrating only on checking $CPT$ symmetry in the mass matrix
of the $B_{d}$ mesons, here we assume that semileptonic $B$ decays
satisfy the $\Delta Q=\Delta B$ rule and $CPT$ invariance. We also
neglect $CPT$ violation in the transition amplitudes for nonleptonic
$B$ decays. Relaxing these limitations can certainly be done, but the
results will become too complicated. To obtain simple and instructive results,
we further assume that
\begin{equation}
\frac{\Delta \Gamma}{\Gamma}\; =\; 0\; ,\;\;\;\;\;
{\rm Im}\phi\; =\; 0\; ,\;\;\;\;\;
{\rm Im}\theta\; =\; 0\; ,
\end{equation}
and ${\cal S}=\cot\theta$ is small and real. As examined in Ref.[3],
large ${\rm Im}\phi$ and $|{\cal S}|$ should be measured in the
dileptonic decay ratios of $B^{0}_{d}\bar{B}^{0}_{d}$ pairs at the $\Upsilon
(4S)$.
On the other hand, a mode-independent analysis shows that
$\Delta \Gamma/\Gamma$ is of the order $O(\leq 10^{-2})$ and has little
effect on the time-integrated $CP$ asymmetries in $B_{d}$ decays [7].
In this work, we proceed with the above approximations to calculate the decay
probabilities of $B^{0}_{d}$ and $\bar{B}^{0}_{d}$ mesons and
explore the $CPT$-violating effects on $CP$
asymmetries instructively. The more precise results
will be presented elsewhere.

	In view of current experiments on $B$-meson physics and proposals
for future $B$ factories [8], we consider two categories of experimental
environments for the production and decays of neutral $B$ mesons.

\begin{center}
{\bf A. $~$ Decays of incoherent $B^{0}_{d}$ and $\bar{B}^{0}_{d}$
mesons}
\end{center}

	In a hadronic production environment [9] or in high energy
$e^{+}e^{-}$ reactions (e.g., at the $Z$ peak), $B^{0}_{d}$ and
$\bar{B}^{0}_{d}$
mesons are produced incoherently. The identification of the flavor of
a neutral $B$ meson can make use of hadrons produced nearby in phase
space [10]. In this case, the time-dependent probability for $B^{0}_{d}$
($\bar{B}^{0}_{d}$)
decaying into a hadronic state $|f>$ is given by
\begin{eqnarray}
{\rm Pr}[B^{0}_{d}(t)\rightarrow f] & \propto & |A_{f}|^{2}e^{-\Gamma t}
|g_{+}(t)+\tilde{g}_{+}(t)\zeta_{f}|^{2} \; , \nonumber\\
{\rm Pr}[\bar{B}^{0}_{d}(t)\rightarrow f] & \propto & |A_{f}|^{2}
e^{-\Gamma t}|\tilde{g}_{-}(t)+g_{-}(t)\zeta_{f}|^{2} \; ,
\end{eqnarray}
where
\begin{equation}
A_{f}\; \equiv \; <f|H|B^{0}_{d}>\; , \;\;\;\;\; \bar{A}_{f}\;
\equiv \; <f|H|\bar{B}^{0}_{d}>\; , \;\;\;\;\; \zeta_{f}\;
\equiv \; \frac{\bar{A}_{f}}{A_{f}}\; .
\end{equation}
With the approximations made in Eq.(5), we simplify Eq.(6) as
\begin{eqnarray}
{\rm Pr}[\stackrel{(-)}{B^{0}_{d}}(t)\rightarrow f] & \propto &
|A_{f}|^{2}e^{-\Gamma t}\left \{ (1+|\xi_{f}|^{2})\stackrel{(-)}{+}
(1-|\xi_{f}|^{2})\cos (\Delta m t) \right . \nonumber\\
&   & \left . \stackrel{(+)}{-}2{\rm Im}\xi_{f}\sin (\Delta m t)
\stackrel{(-)}{+}2{\cal S}{\rm Re}\xi_{f}[1-\cos (\Delta m t)]\right \} \; ,
\end{eqnarray}
where $\xi_{f}\equiv e^{i\phi}\zeta_{f}$, $|\xi_{f}|=|\zeta_{f}|$,
and those $O({\cal S}^{2})$
terms have been neglected.
Integrating ${\rm Pr}[\stackrel{(-)}{B^{0}_{d}}(t)\rightarrow f]$ over
$t$, we obtain the time-independent decay probabilities:
\begin{equation}
{\rm Pr}(\stackrel{(-)}{B^{0}_{d}}\rightarrow f) \; \propto \;
|A_{f}|^{2} \left [\frac{1+|\xi_{f}|^{2}}{2}\stackrel{(-)}{+}
\frac{1}{1+x^{2}_{d}}\frac{1-|\xi_{f}|^{2}}{2}
\stackrel{(+)}{-}\frac{x_{d}}{1+x^{2}_{d}}{\rm Im}\xi_{f}
\stackrel{(-)}{+}\frac{x^{2}_{d}}{1+x^{2}_{d}}{\cal S}{\rm Re}\xi_{f}
\right ] \; ,
\end{equation}
where $x_{d}\equiv \Delta m/\Gamma$ is a measurable for
$B^{0}_{d}-\bar{B}^{0}_{d}$
mixing. Different from the previous results in the literature,
here a linear ${\cal S}$ term, which implies $CPT$ violation in
the $B^{0}_{d}\bar{B}^{0}_{d}$ mass matrix, appears in the decay probabilities.
Note that in Ref.[3] $|\zeta_{f}|=1$ has been taken for
$B^{0}_{d}/\bar{B}^{0}_{d}\rightarrow \psi K_{S}$. Such an approximation
is generally inappropriate (see Sect.III.A), since significant direct
$CP$ violation may exist in decay amplitdes for many neutral $B$ channels.

	For the case that $B^{0}_{d}$ ($\bar{B}^{0}_{d}$) decays into $|\bar{f}>$,
the $CP$-conjugate state of $|f>$, the corresponding decay probability
${\rm Pr}[B^{0}_{d}(t)\rightarrow \bar{f}]$
(${\rm Pr}[\bar{B}^{0}_{d}(t)\rightarrow \bar{f}]$) can be obtained from
${\rm Pr}[\bar{B}^{0}_{d}(t)\rightarrow f]$ (${\rm Pr}[B^{0}_{d}(t)\rightarrow
f]$)
by the replacements
$A_{f}\rightarrow \bar{A}_{\bar{f}}$,
$\zeta_{f}\rightarrow \bar{\zeta}_{\bar{f}}$, and
$\xi_{f}\rightarrow \bar{\xi}_{\bar{f}}$ , where
\begin{equation}
\bar{A}_{\bar{f}}\; \equiv \; <\bar{f}|H|\bar{B}^{0}_{d}> \; ,
\;\;\;\; A_{\bar{f}}\; \equiv \; <\bar{f}|H|B^{0}_{d}> \; , \;\;\;\;
\bar{\zeta}_{\bar{f}} \; \equiv \; \frac{A_{\bar{f}}}{\bar{A}_{\bar{f}}} \; ,
\;\;\;\; \bar{\xi}_{\bar{f}}\; \equiv \; e^{-i\phi}\bar{\zeta}_{\bar{f}}\; .
\end{equation}
With the same replacements, one can obtain the time-integrated decay
probabilities ${\rm Pr}(\stackrel{(-)}{B^{0}_{d}}\rightarrow
\bar{f})$ from Eq.(9).

\begin{center}
{\bf B. $~$ Decays of coherent $B^{0}_{d}\bar{B}^{0}_{d}$
pairs}
\end{center}

	At the $\Upsilon (4S)$ resonance, the $B$'s are produced in a two-body
$(B^{+}_{u}B^{-}_{u}$ and $B^{0}_{d}\bar{B}^{0}_{d}$) state with definite
charge parity. The two neutral $B$ mesons mix coherently until one of them
decays. Thus one can use the semileptonic decays of one meson to tag the
flavor of the other meson decaying into a flavor-nonspecific hadronic
final state.
The unique experimental advantages of studying $b$-quark physics
in this energy region are well known [11], and both symmetric and asymmetric
$e^{+}e^{-}$ $B$ factories will be built in the near future on the basis of
such
threshold collisions [12].

	The time-dependent wave function for a
$B^{0}_{d}\bar{B}^{0}_{d}$ pair produced at the $\Upsilon (4S)$
can be written as
\begin{equation}
\frac{1}{\sqrt{2}}\left [|B^{0}_{d}(\vec{k},t)>\otimes
|\bar{B}^{0}_{d}(-\vec{k},t)> +C|B^{0}_{d}(-\vec{k},t)>\otimes
|\bar{B}^{0}_{d}(\vec{k},t)>\right ] \; ,
\end{equation}
where $\vec{k}$ is the three-momentum vector of the $B_{d}$ mesons,
and $C=\pm$ is the charge parity of the $B^{0}_{d}\bar{B}^{0}_{d}$ pair.
Supposing one $B_{d}$ meson decaying into a
semileptonic state $|l^{\pm}X^{\mp}>$ at (proper) time $t_{1}$ and
the other into a nonleptonic state $|f>$ at time $t_{2}$, the time-dependent
probabilities for such joint decays are given by
\begin{eqnarray}
{\rm Pr}(l^{+}X^{-},t_{1};f,t_{2})_{C} & \propto & |A_{l}|^{2}
|A_{f}|^{2}e^{-\Gamma (t_{1}+t_{2})} \left | g_{+}(t_{1})[
\tilde{g}_{-}(t_{2}) + \zeta_{f}g_{-}(t_{2})] \right . \nonumber\\
&   & \left . +C\tilde{g}_{-}(t_{1})[g_{+}(t_{2})+\zeta_{f}\tilde{g}_{+}
(t_{2})]\right |^{2} \; , \nonumber\\
{\rm Pr}(l^{-}X^{+},t_{1};f,t_{2})_{C} & \propto & |A_{l}|^{2}
|A_{f}|^{2}e^{-\Gamma (t_{1}+t_{2})} \left | \tilde{g}_{+}(t_{1})[
\tilde{g}_{-}(t_{2})+\zeta_{f}g_{-}(t_{2})] \right . \\
&   & \left . +Cg_{-}(t_{1})[g_{+}(t_{2}) + \zeta_{f}\tilde{g}_{+}(t_{2})]
\right |^{2} \; , \nonumber
\end{eqnarray}
where $A_{l}\equiv <l^{+}X^{-}|H|B^{0}_{d}> \stackrel{CPT}{==}
<l^{-}X^{+}|H|\bar{B}^{0}_{d}>$.
By applying Eq.(5), Eq.(6) is further simplified as
\begin{eqnarray}
{\rm Pr}(l^{\pm}X^{\mp}, t_{1}; f, t_{2})_{C} & \propto &
|A_{l}|^{2} |A_{f}|^{2} e^{-\Gamma (t_{1}+t_{2})}
\left \{ (1+|\xi_{f}|^{2}) \mp (1-|\xi_{f}|^{2})
\cos [\Delta m (t_{2}+Ct_{1})] \right . \nonumber\\
&   & \pm 2{\rm Im}\xi_{f}\sin [\Delta m (t_{2}+Ct_{1})] \\
&   & \left . \pm 2{\cal S}{\rm Re}\xi_{f} \left [ C-(1+C)\cos( \Delta m t)
+\cos [\Delta m (t_{2}+Ct_{1})] \right ] \right \} \; , \nonumber
\end{eqnarray}
Similar to Eq.(8), here a $CPT$-violating term proportional to
${\cal S}$ appears.

	Within limits of our present detector technology, we have to
consider the feasibility for an $e^{+}e^{-}$ collider to measure the
time development of the decay probabilities
and $CP$ asymmetries. For a symmetric collider running at the
$\Upsilon (4S)$ resonance, the mean decay lenth of $B$'s is
insufficient for the measurement of
$(t_{2}-t_{1})$ [11]. On the other hand, the quantity $(t_{2}+t_{1})$ cannot
be measured in a symmetric or asymmetric storage ring operating at the
$\Upsilon (4S)$, unless the bunch lengths are much shorter than the decay
lengths [11,12]. Therefore, only the time-integrated measurements are available
at
a symmetric $B$ factory. Integrating ${\rm Pr}(l^{\pm}X^{\mp},t_{1};
f,t_{2})_{C}$ over $t_{1}$ and $t_{2}$, we obtain
\begin{equation}
{\rm Pr}(l^{\pm}X^{\mp},f)_{-} \; \propto \; |A_{l}|^{2}|A_{f}|^{2}\left [
\frac{1+|\xi_{f}|^{2}}{2}\mp \frac{1}{1+x^{2}_{d}}
\frac{1-|\xi_{f}|^{2}}{2} \mp \frac{x^{2}_{d}}{1+x^{2}_{d}}
{\cal S}{\rm Re}\xi_{f}\right ] \; ,
\end{equation}
and
\begin{eqnarray}
{\rm Pr}(l^{\pm}X^{\mp},f)_{+} & \propto & |A_{l}|^{2}|A_{f}|
^{2} \left [\frac{1+|\xi_{f}|^{2}}{2}\mp \frac{1-x^{2}_{d}}
{(1+x^{2}_{d})^{2}}\frac{1-|\xi_{f}|^{2}}{2}
\pm\frac{2x_{d}}{(1+x^{2}_{d})^{2}}{\rm Im}\xi_{f} \right . \nonumber\\
&   & \left . \mp \frac{x^{2}_{d}(1-x^{2}_{d})}{(1+x^{2}_{d})^{2}}
{\cal S}{\rm Re}\xi_{f}\right ] \; .
\end{eqnarray}
For an asymmetric collider running in this energy region, one might want to
integrate Eq.(13) only over $(t_{2}+t_{1})$ in order to measure the
development of the decay probabilities with $\Delta t\equiv (t_{2}-t_{1})$
[11].
In this case, we obtain
\begin{eqnarray}
{\rm Pr}(l^{\pm}X^{\mp},f;\Delta t)_{-} & \propto & |A_{l}|^{2}|A_{f}|^{2}
e^{-\Gamma |\Delta t|}\left \{ \frac{1+|\xi_{f}|^{2}}{2}
\mp \frac{1-|\xi_{f}|^{2}}{2}\cos (\Delta m \Delta t) \right . \nonumber\\
&  & \left . \pm {\rm Im}\xi_{f}\sin (\Delta m \Delta t)
\mp {\cal S}{\rm Re}\xi_{f} [1-\cos (\Delta m \Delta t)]\right \} \; ,
\end{eqnarray}
and
\begin{eqnarray}
{\rm Pr}(l^{\pm}X^{\mp},f;\Delta t)_{+} & \propto & |A_{l}|^{2}|A_{f}|^{2}
e^{-\Gamma |\Delta t|} \left \{ \frac{1+|\xi_{f}|^{2}}{2} \mp\frac{1}
{\sqrt{1+x^{2}_{d}}}\frac{1-|\xi_{f}|^{2}}{2}\cos (\Delta m \Delta t +
\phi_{x_{d}}) \right . \nonumber\\
&  & \left . \pm \frac{1}{\sqrt{1+x^{2}_{d}}}{\rm Im}\xi_{f}\sin
(\Delta m \Delta t +\phi_{x_{d}}) \right . \\
&   & \left . \mp {\cal S}{\rm Re}\xi_{f} \left
[\frac{4-x^{2}_{d}}{4+x^{2}_{d}}-
\frac{1}{\sqrt{1+x^{2}_{d}}}\cos (\Delta m \Delta t +\phi_{x_{d}})
\right ] \right \} \; , \nonumber
\end{eqnarray}
where $\phi_{x_{d}}\equiv \tan^{-1}x_{d}$.

	For the case that one neutral $B$ meson decays into $|l^{\mp}X^{\pm}>$ at time
$t_{1}$ and the other decays
into $|\bar{f}>$ (the $CP$-conjugate state of $|f>$) at time $t_{2}$, the
corresponding decay probabilities
${\rm Pr}(l^{\mp}X^{\pm},t_{1};\bar{f},t_{2})_{C}$
can be obtained from Eq.(13) by the replacements in Eq.(10).
In a similar manner, one can obtain ${\rm Pr}(l^{\mp}X^{\pm},\bar{f})_{\pm}$
and ${\rm Pr}(l^{\mp}X^{\pm},\bar{f};\Delta t)_{\pm}$
straightforwardly from Eqs.(14-17). \\

\begin{flushleft}
{\Large\bf III. $~$ $CP$ asymmetries and $CPT$-violating effects}
\end{flushleft}

	The difference between the decay probabilities associated with
$B^{0}_{d}\rightarrow f$ and $\bar{B}^{0}_{d}\rightarrow \bar{f}$ is
a basic signal for $CP$ violation. For the decays of incoherent
$B^{0}_{d}$ and $\bar{B}^{0}_{d}$ mesons, the time-dependent and
time-integrated $CP$-violating asymmetries are defined by
\begin{equation}
{\cal A}(t) \; \equiv \; \frac{{\rm Pr}[B^{0}_{d}(t)\rightarrow f]-
{\rm Pr}[\bar{B}^{0}_{d}(t)\rightarrow \bar{f}]}
{{\rm Pr}[B^{0}_{d}(t)\rightarrow f]+{\rm Pr}[\bar{B}^{0}_{d}(t)
\rightarrow \bar{f}]} \;
\end{equation}
and
\begin{equation}
{\cal A} \; \equiv \; \frac{{\rm Pr}(B^{0}_{d}\rightarrow f)-{\rm Pr}
(\bar{B}^{0}_{d}\rightarrow \bar{f})}
{{\rm Pr}(B^{0}_{d}\rightarrow f)+{\rm Pr}(\bar{B}^{0}_{d}\rightarrow
\bar{f})} \; ,
\end{equation}
respectively.
Corresponding to the possible measurements for
joint $B^{0}_{d}\bar{B}^{0}_{d}$ decays at symmetric (S) and
asymmetric (A) $e^{+}e^{-}$ $B$ factories, we define the $CP$-violating
asymmetries as
\begin{equation}
{\cal A}^{\rm S}_{C} \; \equiv \; \frac{{\rm Pr} (l^{-}X^{+},f)_{C}
-{\rm Pr} (l^{+}X^{-},\bar{f})_{C}}
{{\rm Pr} (l^{-}X^{+},f)_{C}+{\rm Pr} (l^{+}X^{-},\bar{f})_{C}} \; ,
\end{equation}
and
\begin{equation}
{\cal A}^{\rm A}_{C}(\Delta t) \; \equiv \; \frac{{\rm Pr}(l^{-}X^{+},f;\Delta
t)
_{C}-{\rm Pr}(l^{+}X^{-},\bar{f};\Delta t)_{C}}
{{\rm Pr}(l^{-}X^{+}, f;\Delta t)_{C}+{\rm Pr}(l^{+}X^{-}, \bar{f};\Delta t)
_{C}} \; .
\end{equation}
In the following, we calculate these asymmetries for two categories of neutral
$B$
decays and discuss the small $CPT$-violating effects on them.

\begin{center}
{\bf A. $~$ $CP$-eigenstate decays}
\end{center}

	We first consider the $B^{0}_{d}$ and $\bar{B}^{0}_{d}$ decays
to $CP$ eigenstates (i.e., $|\bar{f}>=\pm |f>$)
such as $\psi K_{S}, \pi^{+}\pi^{-}$, and $\pi^{0}K_{S}$.
With the phase convention $CP|B^{0}_{d}>=|\bar{B}^{0}_{d}>$ and the
approximations in Eq.(5), we
have $A_{\bar{f}}=\pm A_{f}, \bar{A}_{\bar{f}}=\pm \bar{A}_{f}$,
$\bar{\zeta}_{\bar{f}}=1/\zeta_{f}$, and $\bar{\xi}_{\bar{f}}=1/\xi_{f}$.
For convenience, we define three characteristic quantities:
\begin{equation}
{\cal U}\; =\; \frac{1-|\xi_{f}|^{2}}{1+|\xi_{f}|^{2}}\; ,
\;\;\;\;\; {\cal V}\; =\; \frac{-2{\rm Im}\xi_{f}}{1+|\xi_{f}|^{2}}\; ,
\;\;\;\;\; {\cal W}\; =\; \frac{2{\cal S}{\rm Re}\xi_{f}}{1+|\xi_{f}|^{2}} \; .
\end{equation}
Nonvanishing ${\cal U}$ and ${\cal V}$ imply the $CP$ violation
in the decay amplitude and the one from interference between decay and
mixing [13], respectively. ${\cal W}$ (proportional to ${\cal S}$)
is a measure of $CPT$ violation in the
mass matrix of the neutral $B$ mesons.

	For the decays of incoherent $B^{0}_{d}$ and $\bar{B}^{0}_{d}$ mesons into
$CP$ eigenstates, we obtain the $CP$ asymmetries:
\begin{equation}
{\cal A}(t) \; = \; {\cal U}\cos (\Delta m t)
+{\cal V}\sin (\Delta m t)+{\cal W}[1-\cos (\Delta m t)] \; ,
\end{equation}
and
\begin{equation}
{\cal A} \; = \; \frac{1}{1+x^{2}_{d}}{\cal U}
+\frac{x_{d}}{1+x^{2}_{d}}{\cal V}
+\frac{x^{2}_{d}}{1+x^{2}_{d}}{\cal W} \; .
\end{equation}
For symmetric and asymmetric $e^{+}e^{-}$ collisions
at the $\Upsilon (4S)$, the
corresponding $CP$ asymmetries in $(B^{0}_{d}\bar{B}^{0}_{d})_{C}
\rightarrow (l^{\pm}X^{\mp})f$ are given by
\begin{eqnarray}
{\cal A}^{\rm S}_{-} & = & \frac{1}{1+x^{2}_{d}}{\cal U}
+\frac{x^{2}_{d}}{1+x^{2}_{d}}
{\cal W} \; , \nonumber \\
{\cal A}^{\rm S}_{+} & = & \frac{1-x^{2}_{d}}{(1+x^{2}_{d})^{2}}{\cal U}
+\frac{2x_{d}}{(1+x^{2}_{d})^{2}}{\cal
V}+\frac{x^{2}_{d}(1-x^{2}_{d})}{(1+x^{2}_{d})^{2}}
{\cal W} \; ;
\end{eqnarray}
and
\begin{eqnarray}
{\cal A}^{\rm A}_{-}(\Delta t) & = & {\cal U}\cos (\Delta m \Delta t)
+{\cal V}\sin (\Delta m \Delta t)+{\cal W}[1-\cos (\Delta m
\Delta t)] \; , \nonumber \\
{\cal A}^{\rm A}_{+}(\Delta t) & = & \frac{1}{\sqrt{1+x^{2}_{d}}}
\left [{\cal U}\cos (\Delta m \Delta t + \phi_{x_{d}})
+ {\cal V}\sin (\Delta m \Delta t + \phi_{x_{d}}) \right ] \nonumber\\
&   & + {\cal W}
\left [\frac{4-x^{2}_{d}}{4+x^{2}_{d}}-\frac{1}{\sqrt{1+x^{2}_{d}}}
\cos (\Delta m\Delta t +\phi_{x_{d}})\right ] \; .
\end{eqnarray}
{}From the above equations
we observe that all the $CP$ asymmetries get modified if $CPT$ violation is
present. Two remarks are in order.

	(1) ${\cal A}^{\rm S}_{-}$ is neither a
pure measure of direct $CP$ violation in the decay amplitude (i.e.,
$|\xi_{f}|\neq 1$) nor that of small $CPT$ violation in the mass
matrix, but a combination of both of them.
If ${\cal S}$ happens to take the following special value:
\begin{equation}
{\cal S}\; =\; \frac{1}{2x^{2}_{d}}\frac{|\zeta_{f}|^{2}-1}{{\rm Re}
\xi_{f}} \; ,
\end{equation}
the effects of $CP$ and $CPT$ violation will completely cancel out
in ${\cal A}^{\rm S}_{-}$.
In ${\cal A}$ and ${\cal A}^{\rm S}_{+}$, the $CP$ asymmetry from the
interference between decay and mixing is commonly dominant.
A combination of measurements of ${\cal A}^{\rm S}_{-}$ and
${\cal A}$ (or ${\cal A}^{\rm S}_{+}$) can in principle determine
${\cal V}$ or ${\rm Im}\xi_{f}$ unambiguously, as the relations
\begin{equation}
{\cal V}\; =\; \frac{1+x^{2}_{d}}{x_{d}}\left ({\cal A}
-{\cal A}^{\rm S}_{-}\right )\; =\; \frac{(1+x^{2}_{d})^{2}}
{2x_{d}}\left [{\cal A}^{\rm S}_{+}-
\frac{1-x^{2}_{d}}{1+x^{2}_{d}}{\cal A}^{\rm S}_{-}\right ] \;
\end{equation}
hold.

	(2) $CPT$-violating effects can be probed
by measuring the time development of the $CP$ asymmetries
at the $\Upsilon (4S)$ or at a hadronic $B$ factory.
With the help of
\begin{equation}
{\cal A}^{\rm A}_{+}(\Delta t) \; =\; \frac{1}{\sqrt{1+x^{2}_{d}}}
{\cal A}^{\rm A}_{-}\left (\Delta t+\frac{\phi_{x_{d}}}{\Delta m}
\right )+ \left (\frac{4-x^{2}_{d}}{4+x^{2}_{d}}-\frac{1}{\sqrt{1+x^{2}_{d}}}
\right ) {\cal W} \; ,
\end{equation}
the magnitude of ${\cal W}$ or ${\cal S}$ may be definitely limited by
comparing the data on ${\cal A}^{\rm A}_{+}(\Delta t)$
and ${\cal A}^{\rm A}_{-}(\Delta t)$. In addition,
nonvanishing signals of direct $CP$ violation and $CPT$ violation can appear on
some
special points of ${\cal A}(t)$ and ${\cal A}^{\rm A}_{\pm}(\Delta t)$.
For example,
\begin{equation}
{\cal U}\; =\; {\cal A}\left (\frac{2n\pi}{\Delta m}\right )
\; =\; {\cal A}^{\rm A}_{-}\left (\frac{2n\pi}{\Delta m}\right ) \; ,
\end{equation}
and
\begin{eqnarray}
2{\cal W} & = & {\cal A}\left (\frac{2n\pi}{\Delta m}\right )
+{\cal A}\left (\frac{(2n+1)\pi}{\Delta m}\right ) \nonumber\\
& = & {\cal A}^{\rm A}_{-}\left (\frac{2n\pi}{\Delta m}\right )
+{\cal A}^{\rm A}_{-}\left (\frac{(2n+1)\pi}{\Delta m}\right )  \\
& = & \frac{4+x^{2}_{d}}{4-x^{2}_{d}}\left [{\cal A}^{\rm A}_{+}
\left (\frac{(2n+\frac{1}{2})\pi}{\Delta m}-\frac{\phi_{x_{d}}}{\Delta m}
\right )+{\cal A}^{\rm A}_{+}\left (\frac{(2n-\frac{1}{2})\pi}{\Delta m}
-\frac{\phi_{x_{d}}}{\Delta m}\right )\right ] \; , \nonumber
\end{eqnarray}
where $n=0,\pm 1,\pm 2$, and so on.

	It is worthwhile at this point to emphasize that measuring the
time-integrated asymmetry ${\cal A}^{\rm S}_{-}$ for $B^{0}_{d}$ versus
$\bar{B}^{0}_{d}\rightarrow \psi K_{S}$ is not so good to probe
$CPT$ violation in the $B_{d}$ mass matrix. As
discussed in Ref.[14], the so-called hairpin channel may contribute
to these two decay modes. Thus a small deviation of $|\xi_{\psi K_{S}}|$
from unity is possible. With the help of two-loop effective weak hamiltonians
and
factorization approximations, the ratio of hairpin to
tree-level amplitudes is estimated to be as large as $8\%$ [14]. Therefore,
direct $CP$ violation should be taken into account
when we study the fine effect of $CPT$ violation
in $B^{0}_{d}/\bar{B}^{0}_{d}\rightarrow \psi K_{S}$ and other
neutral $B$ decays.

\begin{center}
{\bf B. $~$ Non-$CP$-eigenstate decays}
\end{center}

	Now we consider the case that $B^{0}_{d}$ and $\bar{B}^{0}_{d}$
decay to a common non-$CP$ eigenstate (i.e., $|\bar{f}>\neq \pm |f>$) but
their amplitudes $A_{f}$ ($A_{\bar{f}}$) and $\bar{A}_{\bar{f}}$
($\bar{A}_{f}$) contain only a single weak phase.
Most of such decays occur through the quark transitions
$\stackrel{(-)}{b}\rightarrow u\bar{c}\stackrel{(-)}{q}$ and
$c\bar{u}\stackrel{(-)}{q}$ (with $q=d,s$), and typical
examples are $B^{0}_{d}/\bar{B}^{0}_{d}\rightarrow
D^{\pm}\pi^{\mp}$ and $\stackrel{(-)}{D}$$^{(*)0}K_{S}$.
In this case, no measurable direct
$CP$ violation arises in the decay amplitudes since $|\bar{A}_{\bar{f}}|
=|A_{f}|, |\bar{A}_{f}|=|A_{\bar{f}}|$,
$|\bar{\zeta}_{\bar{f}}|=|\zeta_{f}|$, and $|\bar{\xi}_{\bar{f}}|=
|\xi_{f}|$ [15]. For convenience, we define
\begin{equation}
\tilde{\cal U} \; =\; \frac{1-|\xi_{f}|^{2}}{1+|\xi_{f}|^{2}}\; ,
\;\;\;\;\; \tilde{\cal V}_{\pm}\; =\; \frac{-{\rm Im}(\xi_{f}\pm
\bar{\xi}_{\bar{f}})}{1+|\xi_{f}|^{2}}\; , \;\;\;\;
\tilde{\cal W}_{\pm}\; =\; \frac{{\cal S}{\rm Re}(\xi_{f}\pm
\bar{\xi}_{\bar{f}})}{1+|\xi_{f}|^{2}} \; .
\end{equation}
Note that here a nonzero $\tilde{\cal U}$ does not mean $CP$
violation in the decay amplitude. $\tilde{\cal V}_{-}$ and
$\tilde{\cal W}_{-}$, which imply indirect $CP$ violation
and $CPT$ violation, will contribute to the $CP$ asymmetries in
this kind of decay modes.

	For the decays of incoherent $B^{0}_{d}$ and $\bar{B}^{0}_{d}$
mesons, we obtain the time-dependent $CP$ asymmetry as follows:
\begin{equation}
{\cal A}(t) \; =\; \frac{\tilde{\cal V}_{-}\sin (\Delta m t)
+\tilde{\cal W}_{-}[1-\cos (\Delta m t)]}
{1+ \tilde{F}(\Delta m t)} \; ,
\end{equation}
where $\tilde{F}$ is a function defined as
\begin{equation}
\tilde{F}(y)\; \equiv \; \tilde{\cal U}\cos y + \tilde{\cal V}_{+}\sin y +
\tilde{\cal W}_{+}(1-\cos y) \; .
\end{equation}
In this case, the time-integrated $CP$ asymmetry is given by
\begin{equation}
{\cal A}\; =\; \frac{x_{d}\tilde{\cal V}_{-}+
x^{2}_{d}\tilde{\cal W}_{-}}
{1+x^{2}_{d}+\tilde{\cal U}+x_{d}\tilde{\cal V}_{+}
+x^{2}_{d}\tilde{\cal W}_{+}} \; .
\end{equation}
For symmetric and asymmetric $e^{+}e^{-}$
collisions at the $\Upsilon (4S)$ resonance, the corresponding
$CP$ asymmetries in the decay modes in question are given as
\begin{eqnarray}
{\cal A}^{\rm S}_{-} & = & \frac{x^{2}_{d}\tilde{\cal W}_{-}}
{1+x^{2}_{d} +\tilde{\cal U} + x^{2}_{d}\tilde{\cal W}_{+}} \; , \nonumber\\
{\cal A}^{\rm S}_{+} & = & \frac{2x_{d}\tilde{\cal V}_{-}
+x^{2}_{d}(1-x^{2}_{d})\tilde{\cal W}_{-}}
{(1+x^{2}_{d})^{2}+(1-x^{2}_{d})\tilde{\cal U}
+2x_{d}\tilde{\cal V}_{+} +x^{2}_{d}(1-x^{2}_{d})\tilde{\cal W}_{+}} \; ;
\end{eqnarray}
and
\begin{eqnarray}
{\cal A}^{\rm A}_{-}(\Delta t) & = & \frac{\tilde{\cal V}_{-}\sin(\Delta
m\Delta t)
+\tilde{\cal W}_{-}[1-\cos (\Delta m\Delta t)]}
{1+ \tilde{F}(\Delta m\Delta t)} \; , \nonumber\\
{\cal A}^{\rm A}_{+}(\Delta t) & = & \frac{\tilde{\cal V}_{-}\sin (\Delta
m\Delta t + \phi_{x_{d}})
+\tilde{\cal W}_{-}\left
[\frac{4-x^{2}_{d}}{4+x^{2}_{d}}\sqrt{1+x^{2}_{d}}-\cos (\Delta m
\Delta t +\phi_{x_{d}})\right ]}
{\sqrt{1+x^{2}_{d}} + \tilde{F}(\Delta m\Delta t+\phi_{x_{d}})
+\tilde{\cal W}_{+}\left
[\frac{4-x^{2}_{d}}{4+x^{2}_{d}}\sqrt{1+x^{2}_{d}}-1\right ]} \; .
\end{eqnarray}

	It should be noted that, in contrast with the asymmetry ${\cal A}^{\rm S}_{-}$
in $B^{0}_{d}\bar{B}^{0}_{d}$ decays into $CP$ eigenstates
(see Eq.(25)), here ${\cal A}^{\rm S}_{-}$ is
a pure measure of $CPT$ violation only if ${\rm Re}\bar{\xi}_{\bar{f}}
\neq {\rm Re}\xi_{f}$.
There is a handful of neutral $B$ decays, e.g.,
$B^{0}_{d}\rightarrow \stackrel{(-)}{D}$$^{0(*)}K_{S},
D^{(*)\pm}\pi^{\mp}$, and $\stackrel{(-)}{D}$$^{(*)0}\pi^{0}$,
which satisfy the conditions $|\bar{\xi}_{\bar{f}}|=|\xi_{f}|$
and ${\rm Re}\bar{\xi}_{\bar{f}}\neq {\rm Re}\xi_{f}$. Taking
$B^{0}_{d}/\bar{B}^{0}_{d}\rightarrow D^{+}\pi^{-}$ for example,
an isospin analysis shows that
\begin{eqnarray}
\xi_{D^{+}\pi^{-}} & = & \frac{V_{cb}V^{*}_{ud}}{V_{cd}V^{*}_{ub}}
\cdot\frac{\bar{a}^{~}_{3/2}e^{i\delta_{3/2}}+\sqrt{2}\bar{a}^{~}_{1/2}
e^{i\delta_{1/2}}}{a^{~}_{3/2}e^{i\delta_{3/2}}-\sqrt{2}a^{~}_{1/2}
e^{i\delta_{1/2}}} \; , \nonumber\\
\bar{\xi}_{D^{-}\pi^{+}} & = & \frac{V_{ud}V^{*}_{cb}}{V_{ub}V^{*}_{cd}}
\cdot\frac{\bar{a}^{~}_{3/2}e^{i\delta_{3/2}}+\sqrt{2}\bar{a}^{~}_{1/2}e^{i\delta_{1/2}}}
{a^{~}_{3/2}e^{i\delta_{3/2}}-\sqrt{2}a^{~}_{1/2}e^{i\delta_{1/2}}}
\; ,
\end{eqnarray}
where $V_{ij}$ ($i=u,c,t; j=d,s,b$) are the Cabibbo-Kobayashi-Maskawa matrix
elements; $a^{~}_{3/2}$ ($\bar{a}^{~}_{3/2}$) and $a^{~}_{1/2}$
($\bar{a}^{~}_{1/2}$) are the isospin amplitudes for a $B^{0}_{d}$
($\bar{B}^{0}_{d}$) decaying into $D^{+}\pi^{-}$ ($D^{-}\pi^{+}$)
and its $CPT$-conjugate process [15], and $\delta_{3/2}$ and $\delta_{1/2}$
are the corresponding strong phases. Obviously $|\bar{\xi}_{D^{-}\pi^{+}}|
=|\xi_{D^{+}\pi^{-}}|$ holds, but $\bar{\xi}_{D^{-}\pi^{+}}\neq
\xi^{*}_{D^{+}\pi^{-}}$ because of $\delta_{3/2}\neq \delta_{1/2}$.
As a result, measurements of ${\cal A}^{\rm S}_{-}$
in such decay modes may serve as a good test of $CPT$
symmetry in the $B$-meson decays.

	From Eqs.(35) and (36) we see that ${\cal A}$ and ${\cal A}^{\rm S}_{+}$ are
also modified
in the presence of nonvanishing $\tilde{\cal W}_{-}$ or ${\cal S}$. However, it
is difficult
to extract any information on $CPT$ violation from them.
In principle, measurements of the time-dependent asymmetries ${\cal A}(t)$ and
${\cal A}^{\rm A}_{\pm}(\Delta t)$ are possible to probe
${\cal S}$ with less ambiguity. For example, nonvanishing $CPT$ violation
can be extracted from
\begin{equation}
{\cal A}\left (\frac{(2n+1)\pi}{\Delta m}\right ) \; =\;
{\cal A}^{\rm A}_{-}\left (\frac{(2n+1)\pi}{\Delta m}\right ) \; = \;
\frac{2\tilde{\cal W}_{-}}
{1-\tilde{\cal U}+2\tilde{\cal W}_{+}} \; ,
\end{equation}
and
\begin{equation}
{\cal A}^{\rm A}_{+}\left (\frac{n\pi}{\Delta m}-\frac{\phi_{x_{d}}}
{\Delta m}\right ) \; = \; \frac{\tilde{\cal W}_{-}
\left [\frac{4-x^{2}_{d}}{4+x^{2}_{d}}\sqrt{1+x^{2}_{d}}-(-1)^{n}\right ]}
{\sqrt{1+x^{2}_{d}} +(-1)^{n}\tilde{\cal U}+ \tilde{\cal W}_{+}
\left [\frac{4-x^{2}_{d}}{4+x^{2}_{d}}\sqrt{1+x^{2}_{d}}-(-1)^{n}\right ]}
\end{equation}
where $n=0,\pm 1,\pm 2$, and so on. \\

\begin{flushleft}
{\Large\bf IV. $~$ Conclusion}
\end{flushleft}

	In order to probe the sources of $CP$ and $CPT$ violation in
neutral $B$-meson decays, we have explored various possible
measurements at $e^{+}e^{-}$ $B$ factories and hadron machines.
It is shown that $CPT$-violating effects can in principle be
distinguished from direct and indirect $CP$-violating effects,
by measuring the time development of $CP$ asymmetries ${\cal A}(t)$ and
${\cal A}^{\rm A}_{\pm}$. A clean signal of $CPT$ violation may appear in
the time-integrated asymmetries ${\cal A}^{\rm S}_{-}$ for some
non-$CP$-eigenstate decay modes, e.g.,
$B^{0}_{d}/\bar{B}^{0}_{d}\rightarrow D^{\pm}\pi^{\mp}$ and
$\stackrel{(-)}{D}$$^{(*)0}K_{S}$. In contrast, measuring
${\cal A}^{\rm S}_{-}$ in the $CP$-eigenstate decays such as
$B^{0}_{d}/\bar{B}^{0}_{d}\rightarrow \psi K_{S}$ and $\pi^{+}\pi^{-}$
cannot provide a good limit on $CPT$ violation, since
direct $CP$ violation may compete with or dominate over $CPT$ violation
in them.

	In keeping with the current interest in tests of
discrete symmetries and conservation laws in the $K$-meson system [16],
the parallel approaches are worth pursuing, especially on
investigating $CP$ violation and checking $CPT$
invariance,
for weak decays of $B$ mesons. With the development in building
high-luminosity $B$ factories [9,11,12], it is possible to observe $CP$
(or $T$) violation in neutral $B$ decays in the near future.
The fine effects of $CPT$ violation, if
they are present, may be probed in the second-round experiments
at a $B$ factory. \\

\begin{flushleft}
{\Large\bf Acknowledgments}
\end{flushleft}

	The author would like to thank Professor H. Fritzsch for his
hospitality and helpful comments on this work. He is also
grateful to Professors A.I. Sanda and D.D. Wu for useful communications.
He finally appreciates the Alexander von Humboldt Foundation
for its financial support.

\newpage

\renewcommand{\baselinestretch}{1.35}


\begin{thebibliography}{99}
\bibitem{15} Here we only consider $CPT$ violation in the $B_{d}$ mass
matrix. Any $CPT$-violating effects in the transition amplitudes of
$B$ decays are neglected.
\end{thebibliography}
\end{document}